\documentclass[preprint2]{aastex}

\usepackage{graphics}
\usepackage{epsfig}

\shorttitle{Gamma-ray excess from stacked blazars}
\shortauthors{J.~Aleksi\'c et al.}

\begin{document}

\title{Gamma-ray excess from a stacked sample of high- and intermediate-frequency peaked blazars observed with the MAGIC telescope}

\author{
J.~Aleksi\'c\altaffilmark{a},
L.~A.~Antonelli\altaffilmark{b},
P.~Antoranz\altaffilmark{c},
M.~Backes\altaffilmark{d},
C.~Baixeras\altaffilmark{e},
J.~A.~Barrio\altaffilmark{f},
D.~Bastieri\altaffilmark{g},
J.~Becerra Gonz\'alez\altaffilmark{h},
W.~Bednarek\altaffilmark{i},
A.~Berdyugin\altaffilmark{j}
% ...
}
\altaffiltext{a} {IFAE, Edifici Cn., Campus UAB, E-08193 Bellaterra, Spain}
\altaffiltext{b} {INAF National Institute for Astrophysics, I-00136 Rome, Italy}
\altaffiltext{c} {Universit\`a  di Siena, and INFN Pisa, I-53100 Siena, Italy}
\altaffiltext{d} {Technische Universit\"at Dortmund, D-44221 Dortmund, Germany}
\altaffiltext{e} {Universitat Aut\`onoma de Barcelona, E-08193 Bellaterra, Spain}
\altaffiltext{f} {Universidad Complutense, E-28040 Madrid, Spain}
\altaffiltext{g} {Universit\`a di Padova and INFN, I-35131 Padova, Italy}
\altaffiltext{h} {Inst. de Astrof\'{\i}sica de Canarias, E-38200 La Laguna, Tenerife, Spain}
\altaffiltext{i} {University of \L\'od\'z, PL-90236 Lodz, Poland}
\altaffiltext{j} {Tuorla Observatory, University of Turku, FI-21500 Piikki\"o, Finland}
% ...


\begin{abstract}

Between 2004 and 2009 a sample of 28 X-ray selected high- and intermediate-frequency peaked blazars with a X-ray flux larger than 2\,$\mu\rm{Jy}$ at 1\,keV in the redshift range from 0.018 to 0.361 was observed with the MAGIC telescope at energies above 100\,GeV. Seven among them were detected and the results of these observations are discussed elsewhere.
Here we concentrate on the remaining 21 blazars which were not detected during this observation campaign 
and present the 3\,sigma (99.7\,\%) confidence upper limits on their flux. The individual flux upper limits lie between 1.6\,\% and 13.6\,\% of the integral flux from the Crab Nebula.
Applying a stacking method to the sample of non-detections with a total of 394.1 hours exposure time, 
we find evidence for an excess with a cumulative significance of 4.9 standard deviations. It is not dominated by individual objects or flares, but increases linearly with the observation time as for a constant source with an integral flux level of $\sim$1.5\,\% of that observed from the Crab Nebula above 150\,GeV.

\end{abstract}

\keywords{BL Lacertae objects: general --- gamma rays: observations}

\section{INTRODUCTION}

MAGIC (\textit{M}ajor \textit{A}tmospheric \textit{G}amma-ray \textit{I}maging \textit{C}herenkov) is currently a system of two 17\,m telescopes located atop the Roque de los Muchachos on the Canary Island of La Palma at 2200\,m a.\,s.\,l. The observations refered to in this study were obtained during the years 2004 - 2009 when MAGIC was still a single-dish telescope. Its 234\,$\rm{m}^2$ tessellated parabolic mirror allows observations of VHE (\textit{V}ery \textit{H}igh \textit{E}nergy) $\gamma$-rays between $\sim$50\,GeV and 10\,TeV.

One key goal of the MAGIC telescope project is to determine the properties of extragalactic VHE sources, among which the high-frequency peaked BL Lacertae objects are the most numerous. 
Blazars are a subclass of radio-loud AGN and belong to the most extreme and powerful objects in the universe. They are characterized by a non-thermal broad-band continuum emission which is highly variable on time scales from years down to minutes \citep{alb2007,aha2007a}.

The spectral energy distribution (SED) of blazars is characterized by two bumps in a $\nu$\,F$_{\nu}$ representation. The first component peaks at energies between IR and hard X-rays, and is assumed to originate from leptonic synchrotron radiation. The maximum of the second peak lies in the $\gamma$-ray energy regime. The origin of this peak can be explained by different and partially concurring models either relying on inverse Compton scattering of electrons \citep{mar,der,sik} or proposing hadronic interactions inside the jet \citep{man,pro}.
In the case the synchrotron peak occurs at energies above $\sim10^{16.5}$\,Hz, \citep[according to][]{nie} these blazars are called HBLs (high-frequency peaked BL Lacertae objects) and for peak energies of $\sim10^{14.5-16.5}$\,Hz IBLs (intermediate BL Lacertae objects).

As of April 2010, altogether 29 blazars were established as VHE sources (24 of them HBLs including M87 as 'misaligned' blazar)\footnote{cf.\ http://wwwmagic.mppmu.mpg.de/$\sim$rwagner/sources/ for an up-to-date list.}, compared to six HBLs, when the MAGIC telescope began its regular observations in December 2004.
The sample presented here comprises 21 X-ray selected objects which were not detected in the VHE regime prior to the MAGIC observations. Nine of the objects were already observed between December 2004 and February 2006 and the upper limits of these observations are reported in \citet{alb2008a}. As there have been improvements within the MAGIC analysis, the data of
these objects were re-analyzed and the new results are presented in this work.
Since no significant detection was attained, upper limits on a 3\,$\sigma$ (99.7\,\%) confidence level will be presented.

None of the observed sources showed any variability on diurnal timescales in the VHE regime. Assuming a positive detection in the case of a flaring state, the observations presented here provide a means of investigating the baseline emission of these objects. 
Therefore, a stacking method applied to the blazar sample can reveal such an emission below the sensitivity limit for each individual object.
Together with VERITAS \citep{ben2009} this is the second stacking analysis which turns out to be successful in the VHE $\gamma$-ray regime. 
Former experiments like HEGRA failed in detecting a significant signal in a stacking analysis due to their limited sensitivity \citep[cf.\ for instance][]{man96}.

In Section \ref{sec:sample} the selection criteria for the objects will be presented. The observations and the data analysis technique are described in Section \ref{sec:obsanal}. The analysis results are shown in Section \ref{sec:results}. Finally, a discussion of the results and inherent implications can be found in Section \ref{sec:disc}.

\section{BLAZAR SAMPLE}\label{sec:sample}

We selected blazars from the compilations from \cite{don} and \cite{CostGhis}. Additionally, some objects were chosen based on the synchrotron peak luminosity from \citet{nie} and one from the sedentary survey by \cite{gio}.

The main selection criteria are the measured X-ray flux at 1\,keV and the distance of the objects. According to \citet{stecker}, the synchrotron flux in the X-ray regime is connected to the flux in the VHE regime by
\begin{equation}
\nu_X F_X \sim \nu_{\rm{TeV}} F_{\rm{TeV}}\label{form:estimation} \ ,
\end{equation}
assuming comparable synchrotron and Compton peak luminosities. Therefore objects with high X-ray fluxes are promising candidates for TeV emission.
As the absorption of $\gamma$-rays within the extragalactic background light \citep[EBL, see e.g.][]{knelow} is energy dependent, it is particularly important in the VHE regime to avoid strong attenuation of $\gamma$-rays by limiting the redshift range. According to \citet{knelow}, at a redshift of $z=0.4$, the expected cutoff energy lies well above 200\,GeV, allowing MAGIC to observe still with its highest sensitivity. Therefore all objects with a maximum redshift $z=0.4$ where considered. The energy threshold of the observations increases with the zenith distance $\theta$. Accounting for this effect, the selection of sources with higher $\theta$ ($30\degr < \theta < 45\degr$)during culmination should be limited to $z < 0.15$. The increasing effect of EBL absorption should, however, imprint itself by a net steepening on the spectrum of the stacked excess.

All criteria are described in detail below. They have been chosen to enhance the probability to detect the sources, hence we selected objects with high fluxes and inverse Compton peaks as well as allowing for the lowest possible energies to be measured with MAGIC.

Compared to \cite{alb2008a}, the selection criteria have been extended. The reason is the enhancement of the sample by taking a wider redshift or zenith distance range into account and including sources whose fitted synchrotron peak flux is high enough even if they show a lower X-ray flux level at 1\,keV.
The sample is divided into four parts:
\begin{itemize}
\item{\bf{I.}  }X-ray selected HBLs obtained from \cite{don} and \cite{CostGhis}: (i) redshift $z < 0.4$, (ii) X-ray flux $F_x(1\,\rm{keV}) > 2\,\mu\rm{Jy}$, and (iii) zenith distance $\theta < 30\degr$ during culmination. Assuming the same luminosities at 1\,keV as at 200\,GeV \citep[following the argumentation of][]{stecker}), the X-ray flux $F_x(1\,\rm{keV}) = 2\,\mu\rm{Jy}$ corresponds to a $\gamma$-ray flux at 200\,GeV of $\sim4.8\cdot 10^{-12}\,\rm{erg}\,\rm{cm^{-2}}\,\rm{s^{-1}}$. This criterion applies to 15 sources including nine sources already observed during cycle 1 of regular MAGIC observations. The sources are listed in Table\ \ref{sample}.
\item{\bf{II.}  }Two HBLs obtained from the same compilations taking a wider range in declination and a lower maximum of the redshift into account: 1ES 0033+59.5 and RXS J1136.5+6737. Selection criteria: (i) redshift $z < 0.15$, (ii) X-ray flux $F_x(1\,\rm{keV}) > 2\,\mu\rm{Jy}$, and (iii) $\theta < 45\degr$ during culmination.
\item{\bf{III.}  }Intermediate BL Lacs taken from \cite{nie} with high peak luminosities at the synchrotron peak. Selection criteria: (i) redshift $z<0.4$, (ii) synchrotron peak frequency $\nu_{\rm{peak}}>2\cdot10^{15}\rm{Hz}$, (iii) flux at the peak $F_{\nu_{\rm{peak}}} > 10^{-11}\,\mbox{erg}\,\mbox{cm}^{-2}\,\mbox{s}^{-1}$, and (iv) zenith angle $\theta<30^\circ$ during culmination. This is valid for three sources: B2 1215+30, PKS 1424+240, and B3 2247+381. All of them can also be found in \cite{don} but with an X-ray flux at 1\,keV below 2\,$\mu\rm{Jy}$. B2 1215+30 is listed there as LBL. In return it is included in the TeV candidate list in \cite{CostGhis}.
\item{\bf{IV.}  }One HBL from the sedentary survey \citep{gio} with the same selection criteria as applied for point one of the sample: 1RXS J044127.8+150455.
\end{itemize}
As several other blazars fulfilling these selection criteria were already detected with MAGIC or other VHE instruments, a post-priori selection was done using only the objects which were not yet detected in the VHE regime in advance of the MAGIC observations leaving 21 objects as discussed herein. All blazars in the MAGIC AGN observation program that fulfill these selection criteria either have been detected (or were known in advance) or they are listed here as non-detections.

Table\ \ref{sample} lists all sources in the sample with relevant parameters. In case of multiple flux or spectral slope measurements the mean value is displayed.

\begin{deluxetable}{lrcccccll}
\footnotesize
\tablewidth{0pt}
\tablecaption{List of targets
\label{sample}} 
\tablehead{
\colhead{Object} &
\colhead{Season} &
\colhead{z} &
\colhead{log($\nu_{p}$)\tablenotemark{a}} &
\colhead{$F_{\nu_p}$\tablenotemark{b}} &
\colhead{$F_X$\tablenotemark{c}} &
\colhead{$\alpha_X$\tablenotemark{c}} & 
\colhead{Cat.\tablenotemark{d}} &
\colhead{Sel.} \\
\colhead{} &
\colhead{} &
\colhead{} &
\colhead{} &
\colhead{} &
\colhead{[$\mu\rm{Jy}$]} &
\colhead{} & 
\colhead{} &
\colhead{crit.\tablenotemark{e}}}
\tablecolumns{10}
\startdata
1ES 0033+595                     & 08/2006 -- 07/2008 & 0.086\tablenotemark{f} & 18.9 & 2.0 & 5.66 &   --   &C$^\dag$, N                    & II\\
1ES 0120+340                     & 08 -- 09/2005      & 0.272                  & 18.3 & 2.5 & 4.34 &  1.93  &C, D$^\dag$, G, N              & I\\
1ES 0229+200\tablenotemark{g}    & 08 -- 11/2006      & 0.140                  & 19.5 & 1.6 & 2.88 &   --   &C$^\dag$, N                    & I,II\\
RX J0319.8+1845\tablenotemark{g} & 12/2004 -- 01/2006 & 0.190                  & 17.0 & 0.4 & 1.76 &  2.07  &D$^\dag$, G, N                 & I\\
1ES 0323+022                     & 09 -- 12/2005      & 0.147                  & 19.9 & 6.3 & 3.24 &  2.46  &C, D$^\dag$, G, N              & I,II\\
1ES 0414+009\tablenotemark{g}    & 12/2005 -- 01/2006 & 0.287     & 20.7 & \hspace{-2mm}10.0 & 5.00 & 2.49  &C, D$^\dag$, G, N              & I\\
1RXS J044127.8+150455            & 10 -- 12/2007      & 0.109                  & --   & --  & 4.74 &  2.10  &G$^\dag$                       & IV\\
1ES 0647+250                     & 02 -- 03/2008      & 0.203\tablenotemark{f} & 18.3 & 3.2 & 6.01 &  2.47  &C$^\dag$, D, N                 & I\\
1ES 0806+524\tablenotemark{g}    & 10 -- 12/2005      & 0.138                  & 16.6 & 1.6 & 4.91 &  2.93  &C, D$^\dag$, N                 & I,II\\
1ES 0927+500                     & 12/2005 -- 02/2006 & 0.188                  & 21.1 & 5.0 & 4.00 &  1.88  &D$^\dag$, G, N                 & I\\
1ES 1011+496\tablenotemark{g}    & 03 -- 04/2006      & 0.212                  & 16.7 & 1.3 & 2.15 &  2.49  &C, D$^\dag$, N                 & I\\
1ES 1028+511                     & 03/2007 -- 02/2008 & 0.361                  & 18.6 & 1.3 & 4.42 &  2.50  &C, D$^\dag$, G, N              & I\\
RGB J1117+202                    & 01/2007 -- 03/2008 & 0.140                  & --   & --  & 6.93 &  1.90  &C$^\dag$, D, G                 & I,II\\
RX J1136.5+6737                  & 02/2007            & 0.135                  & 17.6 & 1.3 & 3.17 &  2.39  &C, D$^\dag$, G, N              & II\\
B2 1215+30                       & 03/2007 -- 03/2008 & 0.237                  & 15.6 & 1.3 & 1.59 &  2.65  &C, D, N$^\dag$\tablenotemark{h}& III\\
2E 1415.6+2557                   & 04/2005 -- 04/2008 & 0.237                  & 19.2 & 3.2 & 3.26 &  2.25  &C, D$^\dag$, G, N              & I\\
PKS 1424+240\tablenotemark{g}    & 05/2006 -- 02/2007 & 0.160\tablenotemark{f} & 15.7 & 1.0 & 1.37 &  2.98  &D, N$^\dag$\tablenotemark{h}   & III\\
RX J1725.0+1152                  & 04/2005 -- 04/2009 & 0.018\tablenotemark{f} & 15.8 & 2.0 & 3.60 &  2.65  &C, D$^\dag$, N                 & I,II\\
1ES 1727+502                     & 05/2006 -- 05/2007 & 0.055                  & 17.4 & 1.3 & 3.36 &  2.61  &C, D$^\dag$, N                 & I,II\\
1ES 1741+196                     & 07/2006 -- 04/2007 & 0.083 & 17.9 & 1.0 & 1.92\tablenotemark{i} &  2.04  &C, D$^\dag$, N                 & I,II\\
B3 2247+381                      & 08 -- 09/2006      & 0.119                  & 15.6 & 1.0 & 0.60 &  2.51  &D, N$^\dag$\tablenotemark{h}   & III\\
\enddata
\tablenotetext{a}{Fitted peak frequency from \citet{nie} in units of log(Hz).}
\tablenotetext{b}{Flux at peak frequency extracted from \citet{nie} in units of $10^{-11}\,\rm{erg}\,\rm{cm}^{-2}\,\rm{s}^{-1}$.}
\tablenotetext{c}{Flux and photon spectral index at 1\,keV.}
\tablenotetext{d}{Compilation where the object appears (C: \cite{CostGhis}; D: \cite{don}; N: \cite{nie}; G: \cite{gio}). The catalog from which the object was selected is marked with a dagger.}
\tablenotetext{e}{Selection criteria which are met by the object.}
\tablenotetext{f}{Tentative redshift.}
\tablenotetext{g}{Known VHE blazar (as of April 2010) due to a detection after the MAGIC observation period.}
\tablenotetext{h}{The objects chosen from \citet{nie} are also listed in \citet{don}, but with a X-ray flux lower than 2\,$\mu\rm{Jy}$.}
\tablenotetext{i}{Mean X-ray flux of multiple measurement in \citet{don} below 2\,$\mu\rm{Jy}$.}
\tablecomments{\,List of objects in the sample of X-ray selected blazars with their observation time windows, redshifts and X-ray measurements.}
\end{deluxetable}

\section{OBSERVATIONS AND DATA ANALYSIS TECHNIQUE}\label{sec:obsanal}

The observations presented here were carried out between December 2004 and April 2009 with a total amount of observation time of 490.0 hours. After quality selection (removing low quality data runs from the analysis) 394.1 hours were used for the analysis or 18.8 hours per source on average. The main reason to discard data from the analysis is a low event rate after image cleaning which is primarily influenced by the weather conditions.

Most of the data were taken in wobble mode. In this mode the pointing position of the telescope is displaced by $0.4\degr$ from the source position. In order to get a well-balanced coverage inside the camera, the wobble position is changed regularly to the opposite (with respect to the source position). Signal and background events are then determined from the same shower images with respect to the source position and to three symmetric OFF positions, respectively, all at the same distance to the camera center. Part of the data of RX J0319.8+1845, 2E 1415.6+2557 and RX J1725.0+1152 were taken in ON mode where the pointing position of the telescope is centered on the object in the sky. For these observations dedicated OFF observations have been used for the background estimation.

The data were processed with the software package MARS \citep{mars} using an automated analysis pipeline.
Details can be found in \citet{bretzwagner2003}, \citet{bretzdorner2008}, and \citet{alb2008b}. Furthermore, the arrival time information of neighboring pixels was taken into account \citep{timing}.

For the separation of signal and background events, dynamic cuts on the distribution of image parameters are applied. The image parameters are moments up to third order in the light distribution of the shower images \citep{Hillas_parameters}. The background suppression is done by means of a parabolic cut in AREA \citep{dyn_cuts} 
and a cut in $\vartheta^2$. The latter parameter is the squared angular distance between the source position and the reconstructed shower origin determined with a refined DISP method \citep{les} taking into account the timing information of the showers.
The $\vartheta^2$-cut used in this analysis is $\vartheta^2<0.0196$ which is a somewhat smaller value than usually used for the Crab Nebula, but provides a better background rejection for weak point sources. The chosen value for $\vartheta^2$ corresponds to a signal region in the camera plane with a diameter of 2.8\ camera pixels. The optical point spread function of the MAGIC telescope during the campaign was smaller than 16.0\,mm corresponding to a diameter of 1.1 pixels, well within this area.
A large sample of objects spanning a long time of observations has to be treated with a robust analysis. The usage of dynamic cuts provides such an analysis on the expense of sensitivity (cf.\ Section\ \ref{sec:Crab}).

The statistical significance for any excess is calculated from the $\vartheta^2$ distribution of signal and background events making use of Eq.\ 17 in \citet{lima}.

Concerning the stacking method as described in Sections\ \ref{sec:stacking} and \ref{sec:spec}, the $\vartheta^2$ distributions have been summed up to retrieve the stacked $\vartheta^2$ signal plot. 
The differential energy spectrum is then calculated from all excess events using average values for the effective collection area and a Monte Carlo correction factor (spill-over correction), weighted each with the exposure time $t_{\rm{exp}}$. The same method has been applied to a data set of the Crab Nebula (cf.\ Section\ \ref{sec:Crab}) demonstrating its feasibility.

\section{RESULTS OF THE MAGIC OBSERVATIONS}\label{sec:results}

During the observation campaign no significant detection of any individual object could be achieved. The results can be found in Table\ \ref{results}. None of the objects showed flaring activity in the VHE band on a significant level on diurnal timescales within the observation time windows. Flaring activity is defined here as an offset of 3 standard deviations from the mean measured $\gamma$-rate for each object. However, flux variations by a factor three would still prevent an individual object of the sample of being detected with high significance.
In this Section we present the upper limits obtained for all 21 objects.

Three of the objects were partially observed during an optical high state within a target of opportunity campaign. The trigger criterion was an increase in the optical flux of the core of more than 50\,\%. The objects are 1ES 0033+595, RGB J1117+202, and B2 1215+30. Significant activity or variability in the VHE $\gamma$-ray regime could not be detected.

\subsection{Crab Nebula Observations\label{sec:Crab}}

\begin{figure}[t]
\begin{center}
\includegraphics[width=0.45\textwidth]{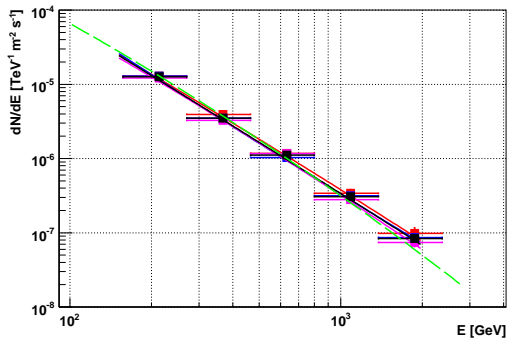}
\end{center}
\caption[]
{\footnotesize
\label{fig:Crab}
Observations of three data sets of the Crab Nebula between October 2005 and January 2008. The red, blue and magenta colored data represent the data sets of the 300\,MHz, 300\,MHz$_{OS}$ and the 2\,GHz systems, respectively. The black curve shows the combined energy spectrum obtained with the stacking method. For comparison the published spectrum from \citet{alb2008c} is plotted as dashed green line. Note that the vertical error bars are hidden by the marks.}
\end{figure}

For a comparative analysis a sample of the Crab Nebula data has been used spanning a time range from Oct 2005 to Jan 2008. Three data sets have been chosen to account for the three different hardware conditions during the blazar observations: 300\,MHz readout system without and with optical splitters and 2\,GHz readout system, later on referred to as 300\,MHz, 300\,MHz$_{OS}$ and 2\,GHz systems, respectively. The $\theta$ distribution of the subsamples have been matched to the one of the blazar sample; the overall observation time after quality selection is $t_{\rm{exp}}=19.2\,\rm{h}$. The individual values as well as the combined result can be found in table\ \ref{Crab}.
The energy spectrum can be fitted with a log parabola \citep[according to Eq.\ 2 in][]{alb2008c} accounting for the flattening of the spectrum towards the inverse Compton peak:
\begin{equation}\label{eq:crab}
\frac{\rm{d}\it{N}}{\rm{d}\it{E}} = f_0 \cdot \left(\frac{E}{300\rm{GeV}}\right)^{\left[a+b\,\rm{log_{10}}(E/300\rm{GeV})\right]}
\end{equation}
with $f_0=(5.37\pm 0.11)\cdot10^{-10}\,\rm{TeV}^{-1}\,\rm{cm}^{-2}\,\rm{s}^{-1}$,\\ $a=-2.20\pm 0.05$ and $b=-0.11\pm 0.03$. 
The $\vartheta^2$ distribution and the energy spectrum have been calculated in the very same way as for the blazar sample by stacking the three individual Crab Nebula samples. 
The integral flux above 150\,GeV is determined to \\$F_{>150\,\rm{GeV}}=(2.81\pm0.05)\cdot 10^{-10}\,\rm{cm}^{-2}\,\rm{s}^{-1}$. It will be used for comparison with the integral upper limits derived from the blazars. Figure\ \ref{fig:Crab} displays the energy spectrum of the stacked excess of the Crab Nebula in comparison to the published spectrum.
The integral flux above 150\,GeV amounts to 91\,\% of that determined in \citet{alb2008c}. A comparison to previous measurements of experiments like HEGRA, H.E.S.S or Whipple is difficult because of the higher energy threshold of these measurements (above 400\,GeV). Due to the hardening of the Crab spectrum towards the peak below 100\,GeV a simple extrapolation of the power-law spectra found there, overestimates the flux at 150\,GeV leading to integral flux ratios of $\sim70--80$\,\% above 150\,GeV \citep[cf.][]{aha2000,aha2006,gru2007}.

\begin{deluxetable}{ccccccccc}
\tablewidth{0pt}
\tablecaption{Observations of the Crab Nebula
\label{Crab}} 
\tablehead{
\colhead{Season} &
\colhead{FADC} &
\colhead{t$_{exp}$} &
\colhead{$\theta$} &
\colhead{Excess} &
\colhead{Backgr.} &
\colhead{Sign.}&
\colhead{$E_{\rm{thr}}$}\\
\colhead{} &
\colhead{system} &
\colhead{[h]} &
\colhead{[\degr]} &
\colhead{events}&
\colhead{events} &
\colhead{$\sigma$} &
\colhead{[GeV]}} 
\startdata
10/2005 - 03/2006 & 300\,MHz        &  3.8 &  6 -- 37 &  967 &  209 & 36.0 & 165\\
09/2006 - 01/2007 & 300\,MHz$_{OS}$ &  8.1 &  7 -- 43 & 2086 &  523 & 51.0 & 165\\
02/2007 - 01/2008 & 2\,GHz          &  7.3 &  8 -- 30 & 2133 &  455 & 53.5 & 165\\
\hline
Combined          & --              & 19.1 &  6 -- 43 & 5188 & 1188 & 82.2 & 165\\
\enddata
\tablecomments{Observations of the Crab Nebula used for a performance test of the stacking method and comparison to the flux upper limits of the blazars. The final spectrum (cf.\ Eq.\ \ref{eq:crab}) is obtained as combination of all the subsamples.}
\end{deluxetable}

The standard MAGIC integral sensitivity is $\sim$\,1.6\,\% of the Crab Nebula flux above 280\,GeV for detecting a signal with 5\,$\sigma$ in 50 hours \citep{timing}. Including lower energies in the integral sensitivity determination, the value increases.
The analysis presented in this work has an integral sensitivity above 150\,GeV of 3.8\,\% of the Crab Nebula flux. This is mainly due to the long-term characteristics of the observations, because the analysis is aimed at a robust and conservative treatment of the data; in addition, also data before the installation of the 2\,GHz system are considered, where the standard MAGIC sensitivity above 280\,GeV is also less with $\sim$\,1.9\,\% of the Crab Nebula flux.

\subsection{Upper Limits}

The upper limits (U.L.) on the excess rates are calculated on a confidence level of 3\,$\sigma$ (99.7 \%) using the method from \citet{rol}. Integral flux upper limits above a given energy are then calculated from them. The integral flux for each source is given above the energy threshold of the analysis, which is defined as the maximum of the differential distribution $\rm{d}\it{N}/\rm{d}\it{E}$ vs $E$ of simulated $\gamma$-showers surviving all cuts. The integral fluxes are also compared to the integral flux of the Crab Nebula above the individual thresholds.

The energy estimation for each source was done based on Monte Carlo simulated $\gamma$ events following a power-law distribution with $\Gamma=-3.0$ for a power law $\rm{d}\it{N}/\rm{d}\it{E}\propto E^\Gamma$. This was done in order to fit better the average spectral slope for the blazars in the VHE regime. For the integral upper limit calculation the same input spectrum ($\Gamma=-3.0$) was used. The resulting upper limits vary between 1.6\,\% and 13.6\,\% of the Crab Nebula flux above the individual energy threshold. 
The energy thresholds lie between 120\,GeV and 230\,GeV, due to differences in the $\theta$ distributions of the individual data samples. The Monte Carlo simulations have been chosen to match exactly the $\theta$ distribution of each data sample. The results of the spectral analysis can be found in Table\ \ref{results}, too. 

\begin{deluxetable}{lrr@{ -- }lrrrrrrc}
\footnotesize
\tablewidth{0pt}
\tablecaption{Results of the analysis
\label{results}}
\tablehead{
\colhead{Object} &
\colhead{$t_{\rm{exp}}$} &
\multicolumn{2}{c}{$\theta$} &
\colhead{Excess} &
\colhead{Backgr.} &
\colhead{Scale} &
\colhead{Sign.} &
\colhead{$E_{\rm{thr}}$\tablenotemark{a}}&
\colhead{U.L.} &
\colhead{U.L.} \\
\colhead{ } &
\colhead{[h]} &
\multicolumn{2}{c}{[\degr]} &
\colhead{events}&
\colhead{events}&
\colhead{} &
\colhead{$\sigma$} &
\colhead{[GeV]} &
\colhead{c.u.\tablenotemark{b}} &
\colhead{f.u.\tablenotemark{c}}}
\tablecolumns{10}
\startdata
1ES 0033+595          &  5.2 & 31 & 41 &   60.0 &  331.0 & 0.33 &   2.8 & 170 &  9.7 & 2.4\\  
1ES 0120+340          & 10.7 &  6 & 18 &   20.7 &  437.3 & 0.33 &   0.9 & 120 &  8.2 & 3.1\\
1ES 0229+200          &  8.0 &  8 & 37 &   55.0 &  572.0 & 0.33 &   2.0 & 120 & 13.6 & 5.1\\
RX J0319.8+1845       & 11.2 & 10 & 31 & --23.4 &  631.4 & 0.59 & --0.7 & 120 &  1.6 & 0.6\\
1ES 0323+022          & 11.4 & 26 & 46 & --45.3 &  751.3 & 0.33 & --1.5 & 170 &  6.9 & 1.7\\
1ES 0414+009          & 18.2 & 28 & 36 &   71.3 & 1020.7 & 0.33 &   1.9 & 170 &  7.7 & 1.9\\
1RXS J044127.8+150455 & 26.9 & 13 & 36 &   18.3 & 1825.7 & 0.33 &   0.4 & 120 &  3.2 & 1.2\\
1ES 0647+250          & 29.2 &  3 & 32 &   64.3 & 1797.7 & 0.33 &   1.3 & 120 &  4.3 & 1.6\\
1ES 0806+524          & 17.5 & 24 & 36 &   17.0 &  752.0 & 0.33 &   0.5 & 140 &  7.2 & 2.2\\
1ES 0927+500          & 16.7 & 21 & 26 &   28.3 &  702.7 & 0.33 &   0.9 & 140 &  5.6 & 1.7\\
1ES 1011+496          & 14.5 & 21 & 29 &   89.0 &  590.0 & 0.33 &   3.1 & 140 &  6.9 & 2.1\\
1ES 1028+511          & 37.1 & 22 & 36 &   65.7 & 2312.3 & 0.33 &   1.2 & 140 &  3.3 & 1.0\\
RGB J1117+202         & 14.9 &  8 & 38 &   25.7 &  804.3 & 0.33 &   0.8 & 120 &  5.3 & 2.0\\
RX J1136.5+6737       & 14.8 & 39 & 46 &   22.7 &  954.3 & 0.33 &   0.6 & 230 &  5.7 & 0.9\\
B2 1215+30            & 16.1 &  1 & 41 &  119.0 &  995.0 & 0.33 &   3.2 & 120 &  9.3 & 3.5\\
2E 1415.6+2557        & 57.4 &  3 & 36 &    7.6 & 3805.4 & 0.54 &   0.1 & 120 &  3.5 & 1.3\\
PKS 1424+240          & 20.0 &  5 & 36 &   51.7 & 1210.3 & 0.33 &   1.3 & 120 &  8.2 & 3.1\\
RX J1725.0+1152       & 32.0 & 17 & 35 &   70.0 & 1859.0 & 0.38 &   1.4 & 140 &  4.2 & 1.3\\
1ES 1727+502          &  6.1 & 21 & 36 &   31.0 &  302.0 & 0.33 &   1.5 & 140 & 11.8 & 3.6\\
1ES 1741+196          & 11.8 &  9 & 40 &   98.7 &  731.3 & 0.33 &   3.1 & 120 &  9.6 & 3.6\\
B3 2247+381           &  8.3 & 10 & 36 &   21.7 &  490.3 & 0.33 &   0.8 & 140 &  5.2 & 1.6\\
\enddata
\tablenotetext{a}{Peak response energy for a power law spectrum $E^\Gamma$ with $\Gamma=-3.0$}
\tablenotetext{b}{Integral flux above $E_{\rm{thr}}$ given in units of the flux of the Crab Nebula (crab units, c.u.)}
\tablenotetext{c}{Integral flux above $E_{\rm{thr}}$ given in flux units f.u. = $10^{-11}\,\rm{cm}^{-2}\,\rm{s}^{-1}$}
\tablecomments{Results of the analysis. The upper limits span a range of 1.6 - 13.6\,\% of the Crab Nebula flux above the corresponding energy threshold.}
\end{deluxetable}

Discovery of VHE $\gamma$-rays from RX J0319.8+1845 and 1ES 0806+524 has recently been reported by the VERITAS collaboration \citep{acc2009,atel0317}, as well as from PKS 1424+240 which was confirmed by the MAGIC collaboration in a campaign independent of the observations presented here \citep{1424arxiv,atel1424M}. The measured VHE flux for the latter source was significantly higher than in previous observations with the MAGIC telescope. 1ES 0229+200 and 1ES 0414+009 have been detected by the H.E.S.S. telescope array in 2006 \citep{aha2007b} and 2009 \citep{atel0414}, respectively. However, since the observations presented here were each performed in advance of the detections mentioned above, the inclusion of these sources in the stacking method is justified. The later detections show that the X-ray selection of possible targets is a reasonable approach.

In order to compare the measured integral fluxes with the upper limits presented here they are extrapolated to the individual energy thresholds as reported in Table\ \ref{results}. 
In all cases except for PKS 1424+240 the upper limits are compatible with the extrapolated reported integral fluxes.

\subsection{Significance Distribution}
Taking a look at the calculated significances of the blazar sample it is evident that most of the individual objects show positive values. Plotting the distribution of the significances, the mean value is not located at 0 as expected for sky regions where no $\gamma$-rays are expected to originate. 

\begin{figure*}[!t]
\begin{center}
\includegraphics[width=0.8\textwidth]{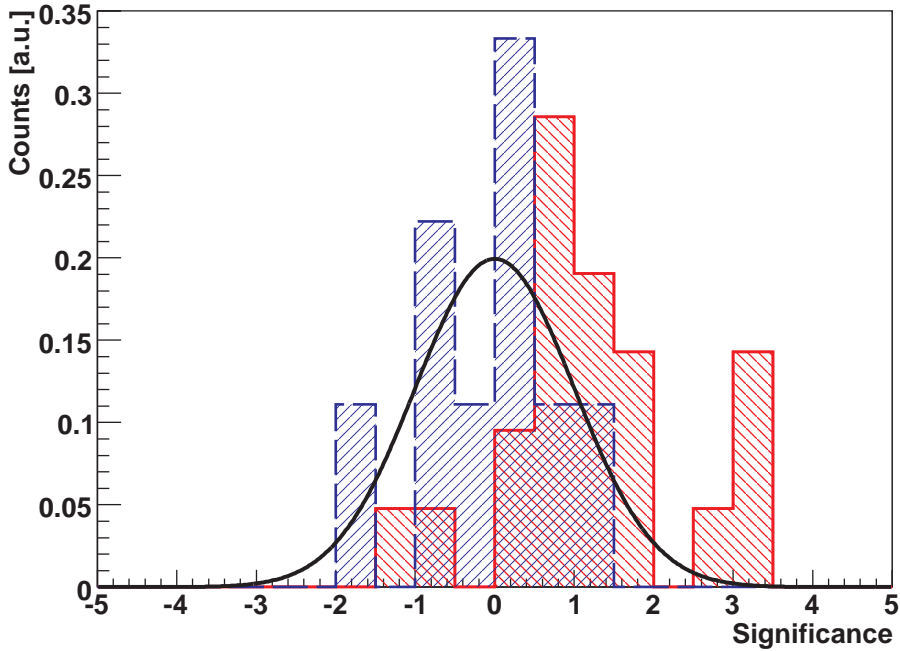}
\end{center}
\caption[]
{\footnotesize
\label{signdistr}
Significance distributions of the blazar (red, hatched up left to low right) and the cross-check sample (blue, hatched low left to up right). The different distributions are normalized to one, so the vertical axis gives the percentage of the whole blazar or cross-check sample, respectively. 
The blazar sample distribution has a mean value of 1.23$\pm$1.17 and the cross-check sample $-$0.08$\pm$0.85.
For comparison a Gaussian with mean value 0 and standard deviation 1 is plotted as black curve.}
\end{figure*}

In Figure\ \ref{signdistr} the significance distribution for the blazar sample is shown together with the result of a cross-check as described below. As the number of individual samples is different for both distributions they have been normalized to one. 
The blazar sample distribution has a mean value of 1.23$\pm$1.17 while the cross-check sample has $-$0.08$\pm$0.85.
This result can be expected due to the fact that our sample is biased by the selection toward potential VHE $\gamma$-ray emitters.

In order to test if the positive signal in the blazar sample originates from a systematic effect of the observations or analysis chain, we also cross-checked this result with data sets obtained as OFF pointings associated to different ON source observations not treated in this paper. These data sets were taken under similar conditions as the blazars covering the whole range of $\theta$ of the blazar sample and processed with the very same analysis chain. The OFF observations were analyzed in wobble mode with respect to two fake source positions in the camera displaced by 0.4\degr from the camera center. Table\ \ref{Off} gives a list of these observations and results. 
Although the fit parameters of Gaussian fits to both distributions do not permit any conclusive statement, a Kolmogorov-Smirnov test of the compatibility of the blazar with the cross-check sample gives a probability of 1.56\,\%. For the Gaussian distributions the test returns 3.42\,\% and 77.03\,\% for the compatibility of the blazar and the cross-check sample with the standard Gaussian, respectively. The cross-check sample is $\sim$\,7 times smaller than the blazar sample, thus systematic effects in the analysis can only largely be ruled out as possible explanation for the shift in the blazar distribution. 

\begin{deluxetable}{crcr@{ -- }lrrr}
\tablewidth{0pt}
\tablecaption{Data samples for cross-check 
\label{Off}} 
\tablehead{
\colhead{Sample} &
\colhead{Season} &
\colhead{$t_{\rm{exp}}$} &
\multicolumn{2}{c}{$\theta$} &
\colhead{Excess} &
\colhead{Backgr.} &
\colhead{Sign.}\\ 
\colhead{} &
\colhead{} &
\colhead{[h]} &
\multicolumn{2}{c}{[\degr]} &
\colhead{events} &
\colhead{events} &
\colhead{$\sigma$}} 
\startdata
1 & 06 -- 07/2006  & 5.4 & 34 & 43 &  --1.3 &  335.3 & --0.1\\
2 & 07/2006        & 3.1 &  6 & 29 &   4.3 &  107.7 &  0.4\\
3 & 11/2006        & 1.9 & 37 & 47 &  19.0 &  255.0 &  1.0\\
4 & 01/2007        & 3.3 & 49 & 56 & --24.7 &  149.7 & --1.8\\
5 & 04/2007        & 2.8 & 11 & 27 &  --9.7 &  139.7 & --0.7\\
6 & 05/2007        & 1.3 & 28 & 37 &   2.0 &   76.0 &  0.2\\
7 & 05/2007        & 7.3 & 29 & 36 & --20.7 &  356.7 & --1.0\\
8 & 01 -- 08/2008  &17.9 & 22 & 38 &   7.0 & 1041.0 &  0.2\\
9 & 02 -- 04/2008  & 9.3 & 22 & 26 &  18.0 &  548.0 &  0.7\\
\enddata
\tablecomments{Data samples used for the cross-check analysis. They were chosen to give a good coverage of the $\theta$ distributions and the different night sky background conditions of the blazar sample.}
\end{deluxetable}

\subsection{Stacking Analysis}\label{sec:stacking}
\begin{figure}[!h]
\begin{center}
\begin{tabular}{c}
\includegraphics[width=0.48\textwidth]{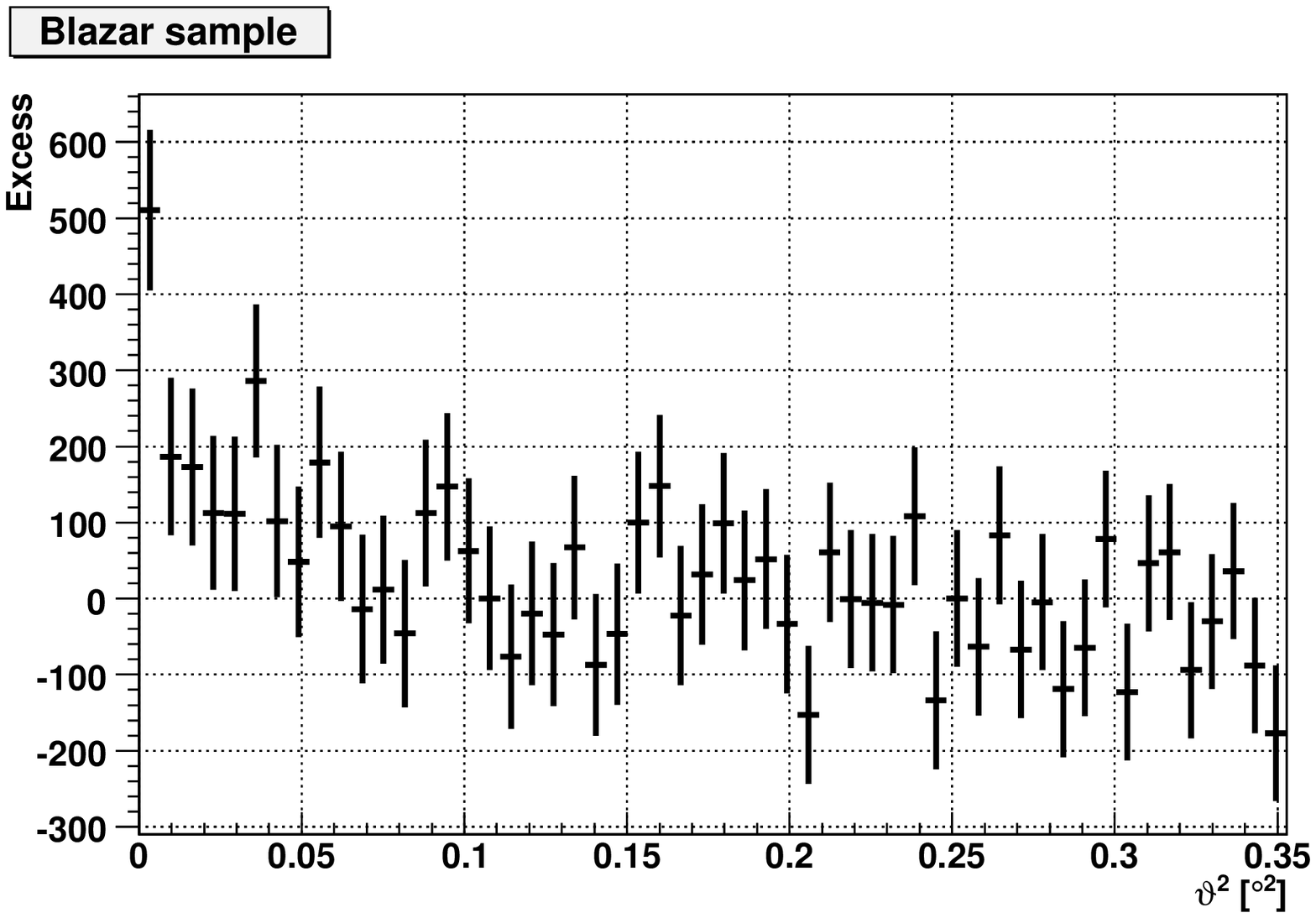}
\\
\includegraphics[width=0.48\textwidth]{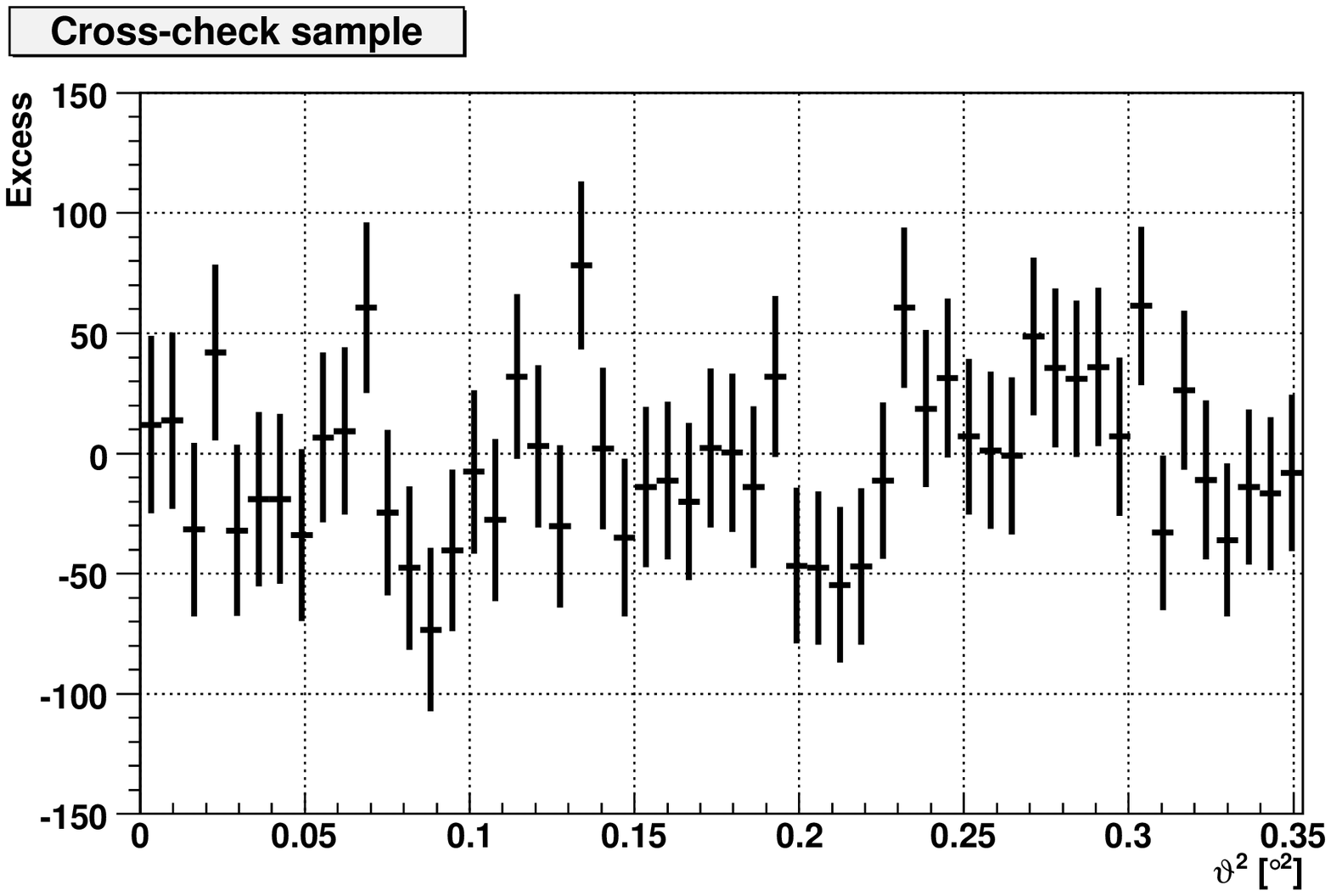}
\\
\end{tabular}
\end{center}
\caption[]
{\footnotesize
\label{stackedtheta}
$\vartheta^2$-distribution of excess events for the stacked blazar sample (top) and the cross-check sample (bottom). The blazar sample shows a clear extension at low values with a significance of 4.9 standard deviations.}
\end{figure}

Even if none of the sources was detected in a single observation, a cumulative signal search seems promising. For this reason the $\vartheta^2$-plots of the individual analyses have been stacked producing one plot for the whole set containing 394.1 hours of data (cf.\ Section\ \ref{sec:obsanal}). Figure\ \ref{stackedtheta} shows the result, a significance of 4.9 standard deviations with 870 excess and 22876 background events. About 30\,\% of the stacked excess comes from blazars now known as VHE $\gamma$-ray emitters. Without these sources the stacked excess amounts to 608 excess events with a significance of 3.8\,$\sigma$ indicating that there are other emitters contained in the sample. Figure\ \ref{fig:exc} underlines this finding.
As expected, the stacked $\vartheta^2$-plot of the cross-check analysis containing no $\gamma$-signal gives a significance of $-$0.1 with $-$6 excess and 3009 background events, the result is shown in Figure\ \ref{stackedtheta} as well.

\subsection{Energy Spectrum}\label{sec:spec}
From the combined excess a differential energy spectrum can be calculated. The differential energy spectrum $dF/dE$ for one source is calculated binwise by dividing the product of the number of excess events N$_{exc,i}$ and the spill over factor a$_i$ by the product of effective collection area A$_{\rm{eff},i}$ and exposure time t$_{exp}$. In order to derive an energy spectrum of the stacked excess, the mean values of a$_i$ and A$_{\rm{eff},i}$ weighted with the observation time have to be taken:
\begin{eqnarray}
\left\langle a_i \right\rangle &=& \frac{\sum_n a_{i,n}\cdot t_{\rm{exp},n}}{\sum_n t_{\rm{exp},n}} \\
\left\langle A_{\rm{eff},i} \right\rangle &=& \frac{\sum_n A_{\rm{eff},i,n}\cdot t_{\rm{exp},n}}{\sum_n t_{\rm{exp},n}}
\end{eqnarray}
with n being the number of objects to be stacked and the energy bin i.
The differential quotient $\rm{d}\it{N}_i/\rm{d}\it{E}$ for each bin can then be calculated as
\begin{equation}
\frac{\rm{d}\it{N}_i}{\rm{d}\it{E}}=\frac{\sum_n N_{\rm{exc},i,n}\cdot \left\langle a_i \right\rangle}{\sum_n t_{\rm{exp},n}\cdot \left\langle A_{\rm{eff},i}\right\rangle\cdot \Delta E_i}\label{eqn:spec}
\end{equation}
with the energy bin width $\Delta$\,E$_i$.
The mean energy spectrum in the observer's frame for all 21 blazars considered in the stacking analysis can be well described by a power law,
\begin{equation}
\frac{\rm{d}\it{N}}{\rm{d}\it{E}} = (2.5 \pm 0.7) \cdot 10^{-11} \frac{1}{\rm{TeV}\rm{cm}^2\rm{s}} \cdot \left(\frac{E}{E_0}\right)^{-3.16 \pm 0.51}
\end{equation}
with $E_0=200\rm{GeV}$.
The differential flux at 200\,GeV corresponds to 1.9\,\% of the one of the Crab Nebula. The integral flux above 150\,GeV is determined as $F=4.3 \cdot 10^{-12}\,\rm{cm}^{-2}\,\rm{s}^{-1}$ corresponding to 1.5\,\% of the integral Crab Nebula flux above 150\,GeV. 

\begin{figure}[t]
\begin{center}
\includegraphics[width=0.48\textwidth]{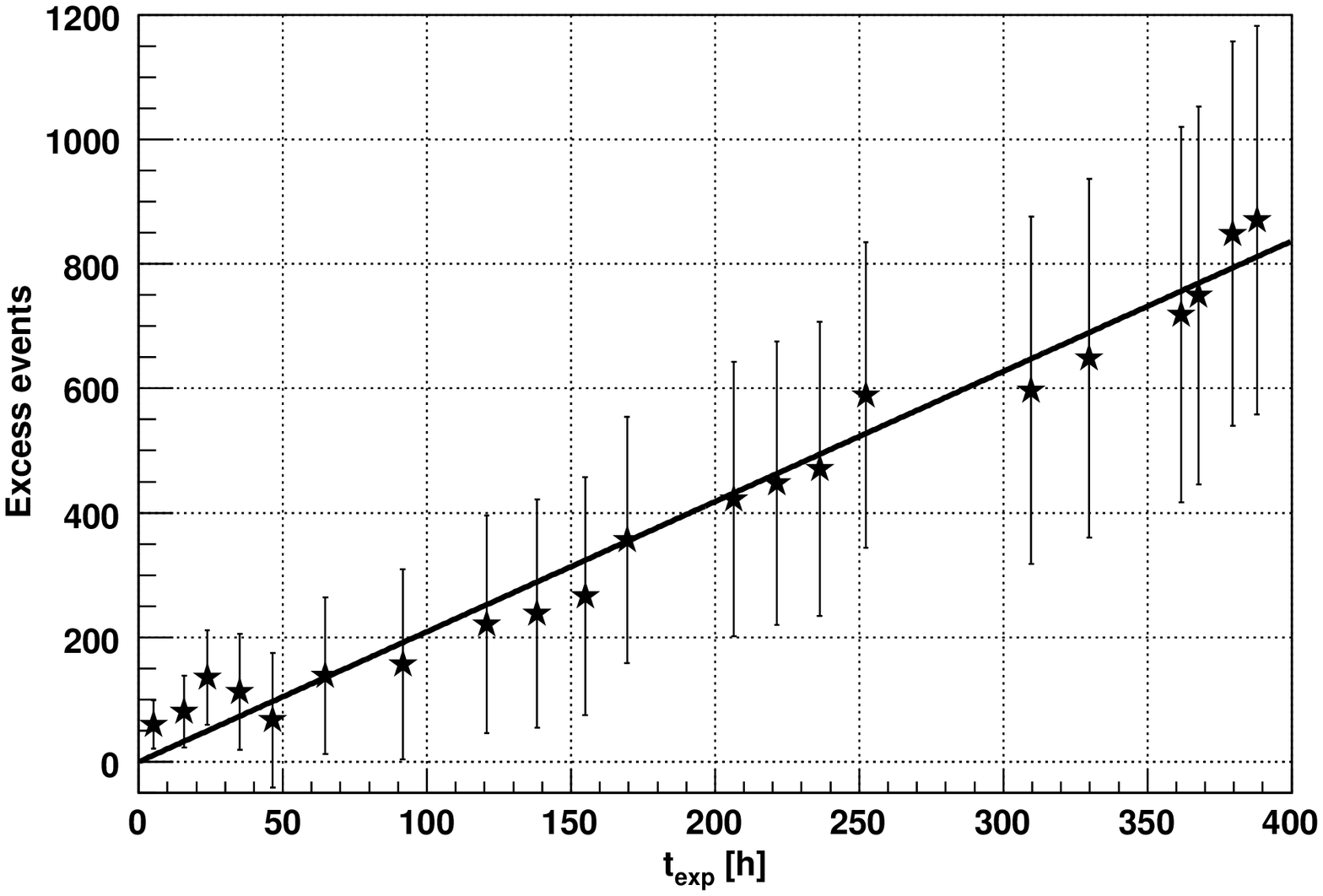}
\end{center}
\caption[]
{\footnotesize
\label{fig:exc}
Excess events of the individual blazars vs. the overall exposure time. On average each blazar contributes with 2.1$\pm$0.3 excess events per hour.}
\end{figure}

On average, each blazar contributes with (2.1$\pm$0.3)/h excess events to the cumulative excess as is illustrated in Figure\ \ref{fig:exc}. The objects are ordered in right ascension.

\begin{figure}[!ht]
\begin{center}
\includegraphics[width=0.48\textwidth]{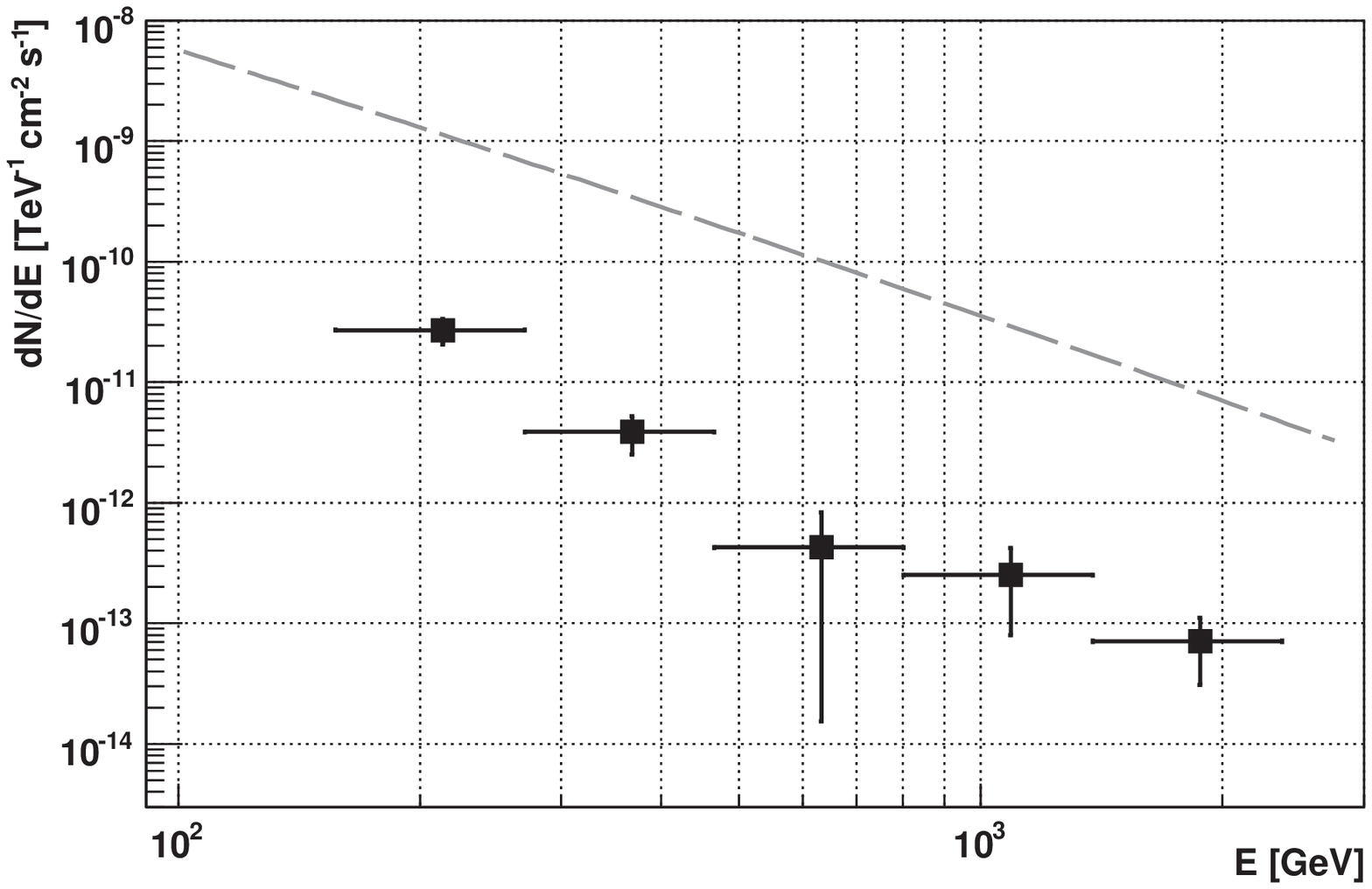}
\end{center}
\caption[]
{\footnotesize
\label{stackedspectrum}
Differential energy spectrum obtained from the stacked source analysis. It is well described by a power law with index $-3.16\pm0.51$. The integral flux above 150\,GeV corresponds to 1.5\,\% of the flux of the Crab Nebula. 
The spectrum of the Crab Nebula is shown as dashed gray line.}
\end{figure}

In Figure\ \ref{stackedspectrum} the measured spectrum is shown.

\section{DISCUSSION}\label{sec:disc}

The positive mean significance distribution indicates that the X-ray selected blazars studied here constitute a fairly representative sample of generic VHE emitters, as suggested by \citet{CostGhis}.
The recent discoveries of individual blazars from the sample indeed corroborate this finding.
The next generation of Cherenkov experiments -- MAGIC-II, H.E.S.S. 2, and later on the Cherenkov Telescope Array 
\citep[CTA,][]{wagner09} 
-- will therefore have good chances to detect an increasing fraction of all known X-ray blazars.

\subsection{Gamma-ray Background}

At 200\,GeV, the attenuation caused by the EBL is negligible, according to the model of \citet{knelow}, so the calculation of the broad-band spectral index $\alpha_{X-\gamma}$ between 1\,keV and 200\,GeV can be done with the observed VHE energy spectrum.
The mean energy flux at 200\,GeV is calculated from the fit to $1.60 \cdot 10^{-12} \rm{erg}\,\rm{cm}^{-2}\,\rm{s}^{-1}$. This value is compared to the 
mean X-ray energy flux at 1\,keV for all sources, weighted with their individual observation time, which is 3.74\,$\mu\rm{Jy}$ corresponding to a flux of $9.05\cdot10^{-12} \rm{erg}\,\rm{cm}^{-2}\,\rm{s}^{-1}$. The ratio of X-ray (1\,keV) to $\gamma$-ray (200\,GeV) flux is
\begin{equation}
\frac{\nu F_{\nu}(1\,\rm{keV})}{\nu F_{\nu}(200\,\rm{GeV})}= 5.66\ ,
\end{equation}
resulting in a broad-band spectral index $\alpha_{X-\gamma}=1.09$.\\
The result suggests that during quiescence the X-ray luminosity is higher than the VHE $\gamma$-ray luminosity above 200\,GeV. Here, we tacitly assume that the X-ray data, which are not contemporaneous with the $\gamma$-ray data, are representative of baseline emission as well. Note, that the X-ray as well as the VHE data are an average over the whole blazar sample considered here and that flux variations commonly observed with the X-ray band do not influence $\alpha_{X-\gamma}$ across eight orders of magnitude. A simple estimation of $\Delta\alpha_{X-\gamma}$ by inferring the error of the average value at 1\,keV of the sample and the error of the energy spectrum at 200\,GeV results in $\Delta\alpha_{X-\gamma}=0.04$. With the newly found X-ray to $\gamma$-ray spectral index of $\alpha_{X-\gamma}=1.09$ one can infer the luminosity function of VHE blazars from their X-ray luminosity function, avoiding the bias toward flares. Assuming equal X-ray and VHE $\gamma$-ray luminosities, HBLs already fail to explain the extragalactic diffuse $\gamma$-ray background \citep{kneiske2008}.

\subsection{Spectral energy distribution}

As no flaring activity has been seen on diurnal scales nor on longer time scales, the cumulative signal of the high-peaked blazars in this sample can be accounted as an upper limit on their baseline emission in VHE $\gamma$-rays, although variability on flux scales below the sensitivity limit of MAGIC may not be excluded.

\begin{figure*}[t]
\begin{center}
\includegraphics[width=0.9\textwidth]{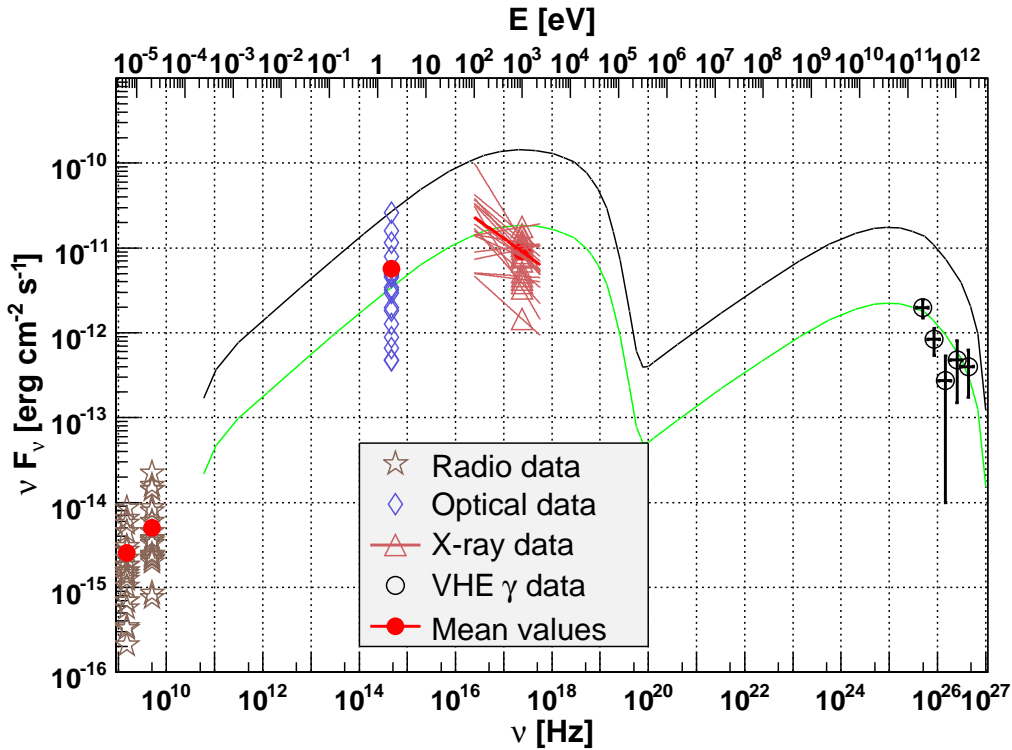}
\end{center}
\caption[]
{\footnotesize
\label{sed}
Spectral energy distribution $\nu F_{\nu}$ vs $\nu$ for the blazar sample. Plotted are the measured fluxes in the radio (1.5\,GHz and 5\,GHz), optical (640\,nm), and X-ray (1\,keV) bands, together with their mean values. In the VHE regime the observed spectrum of the stacked excess is shown. The black curve represents the model fit for 1ES 1959+650 taken from \citet{tal2008} for comparison. The green curve is the same curve scaled down to match the first VHE flux point of the energy spectrum of the stacked excess.}
\end{figure*}

The SED for the blazar sample is determined by taking archival data in the radio and X-ray bands (1.4\,GHz, 5\,GHz, and 1\,keV) if available as well as contemporaneous optical data in the R-Band (640\,nm) taken with the KVA telescope. The collected data are shown in Figure\ \ref{sed}.
In the VHE regime also the deabsorbed spectrum as already shown in Figure\ \ref{stackedspectrum} is displayed. 
From the mean values in optical and X-rays and the mean X-ray spectral index one can infer an average synchrotron peak energy of the sample below 1\,keV.

For a simple comparison the measured spectral energy distribution of the HBL 1ES 1959+650 is drawn. 1ES 1959+650 is a well known VHE blazar which was observed in a historic low emission state in a multiwavelength campaign in 2006 \citep{tal2008}. The differential energy spectrum measured by MAGIC in the VHE regime follows a power law with a photon index $\Gamma=-2.58\pm0.18$. The SED of 1ES 1959+650 can be well fitted with a one-zone synchrotron self-Compton model, which is also plotted in Figure\ \ref{sed}. To guide the eye, the SED is also scaled down to the lowest energy bin of the VHE spectrum of the blazar sample.

\section{CONCLUSIONS}

In the course of the MAGIC observational program during 2004 - 2009, a major part was spent on X-ray bright BL Lacertae objects. 
For 21 non-detections upper limits on the integral flux ranging between 1.6\,\% and 13.6\,\% of the Crab Nebula flux could be determined. 
Applying a stacking method to the individual non-detections we found an average VHE emission of the sample of X-ray selected blazars at the 4.9\,$\sigma$ significance level above 100\,GeV.
It turns out out that the mean VHE $\gamma$-ray flux is significantly lower than in archival X-ray measurements. The two-point spectral index between 1\,keV and 200\,GeV is 1.09$\pm$0.04.

\acknowledgements
We would like to thank the Instituto de Astrofisica de 
Canarias for the excellent working conditions at the 
Observatorio del Roque de los Muchachos in La Palma. 
The support of the German BMBF and MPG, the Italian INFN,
the Swiss National Fund SNF, and the Spanish MICINN is gratefully acknowledged. 
This work was also supported by the Polish MNiSzW Grant N N203 390834, 
by the YIP of the Helmholtz Gemeinschaft, and by grant DO02-353
of the the Bulgarian National Science Fund.

\end{document}